\documentclass[12pt]{elsarticle}
\usepackage[T1]{fontenc}
\usepackage{float}
\usepackage{amsmath}
\usepackage{graphicx}
\usepackage{amssymb}
\biboptions{sort&compress}
\numberwithin{equation}{section}
\numberwithin{figure}{section}

\journal{Nuclear Physics B}
\begin{document}
\begin{frontmatter}
\title{\noindent \textbf{Quantum phase transitions in Bose-Einstein condensates from a Bethe ansatz perspective}}

\author{\textbf{Rubeni, D.$^{1}$; Foerster, A.$^{1}$\footnote[1]{Corresponding author's  e-mail: angela{@}if.ufrgs.br}; Mattei, E.$^{1}$ and Roditi, I.$^{2}$}}
\address{{\small $^{1}$Instituto de F\'{\i}sica da UFRGS, Porto
Alegre, RS - Brazil}\\{\small$^2$Centro Brasileiro de Pesquisas F\'{\i}sicas - CBPF/MCTI, 22290-180, Rio de Janeiro, RJ - Brazil}}
\begin{abstract}
We investigate two solvable models for Bose-Einstein condensates and extract physical information by 
studying the structure of the solutions of their Bethe ansatz equations. A careful observation of these 
solutions for the ground state of both models, as we vary some parameters of the Hamiltonian, suggests a 
connection between the behavior of the roots of the Bethe ansatz equations and the physical behavior of 
the models. Then, by the use of standard techniques for approaching quantum phase transition - gap, 
entanglement and fidelity - we find that the change in the scenery in the roots of the Bethe ansatz equations 
is directly related to a quantum phase transition, thus providing an alternative method for its detection.
\end{abstract}
\end{frontmatter}

\section{Introduction}

The study of Bose-Einstein condensates
 has provided in recent years several new exciting possibilities of 
research, either experimental or theoretical. One of these is the investigation of different phases of 
ultracold atoms. Whether, for instance, one can take them from an insulator to a superfluid behavior, as 
in \cite{Greiner2002}.  Now, given the fact that these condensates reside in the realm of near to 
absolute zero temperatures new tools are needed to approach this quantum phase transition. One possibility 
lies in the field of quantum information and make use of quantities such as the entanglement entropy or the fidelity. 
Another possibility that we would like to advance is the use of quantum integrable systems which contributed with 
several models of Bose-Einstein 
condensates \cite{Links01, jonguan, dukelskyy, ortiz, Kundu, eric5} where one has exact control 
over their solutions through Bethe ansatz methods and whose importance has been highlighted in \cite{hertie, batchelor}. \\

Quantum phase transitions most distinctive feature, when compared with a classical phase transition, is that 
they take place at zero temperature. That is, in practical terms, when  quantum fluctuations are more relevant 
than thermal ones. This is reflected in the behavior of the ground state, of the investigated physical 
system, which changes in some essential manner when some parameter of the associated Hamiltonian takes different values. 
The typical image is presented in \cite{Sachdev2011} for a Hamiltonian given by a linear combination of 
commuting operators, when changing a coefficient of the combination, the ground state may reach a 
non analytic point for some critical value of this coefficient. \\

For quantum integrable systems one can have access to the ground state through the exact solution of the 
corresponding models by the quantum inverse scattering method that leads to the algebraic Bethe ansatz 
equations \cite{Faddeev1979,Faddeev:1996iy,KulSkly,korepin}. A careful observation of the behavior of 
solutions of these equations for the ground state, as we vary some parameters of the 
Hamiltonian, suggests a connection between the behavior of roots of the Bethe ansatz equations and 
the physical behavior of such models. Which is exactly what we expect to happen in quantum phase transitions.\\

Our aim is then, for two different integrable models, the two-site Bose-Hubbard model and the hetero-atomic 
molecular Bose-Einstein condensate (BEC), to compare our results, coming from the study of the solutions of the 
Bethe ansatz equations, with the results of other more standard methods for finding quantum phase transitions 
such as the study of the entanglement, energy gap and fidelity. Notwithstanding that rigorously a quantum phase 
transition is only defined in the limit where N, the number of particles in the system, approaches infinity, the 
previously mentioned concepts show a clear response to variations in the ground state of the system for finite N 
and unequivocally indicate the presence of quantum phase transitions \cite{PhysRevLett.92.212501}. Very recent 
experiments incorporating the realization of a quantum system exhibiting classical bifurcations as well as the 
determination of a suitable criteria for measuring the entanglement between two wells in a 
BEC \cite{PhysRevLett.105.204101, PhysRevLett.106.120404, PhysRevLett.106.120405} turn this kind of investigation 
for such models even more appealing.\\

In the next section we analyze the attractive two-site Bose-Hubbard model where a bifurcation in its 
classical analysis indicates the possibility of a quantum phase transition; we investigate the structure 
of the ground state distribution of roots of the Bethe ansatz equations for this model in order to examine if its 
behavior could be used to unveil a quantum phase transition. A comparison is done with the behavior 
of such quantities as the entanglement, energy gap and fidelity. We also provide an image of which processes occur in the phase transition.
In the third section we apply the same kind of analysis to the hetero-atomic molecular Bose-Einstein 
condensate, which is described by a three mode Hamiltonian that allows for atomic and molecular states 
to exist as a superposition. 
In section 4 we summarize our results and draw some conclusions.

\section{Attractive two-site Bose-Hubbard model}

From the theoretical point of view, the \emph{two-site Bose-Hubbard
model}, also known as the \emph{canonical Josephson Hamiltonian} \cite{RevModPhys.73.307, Links01}, has been a 
useful model in understanding tunneling and self-trapping
phenomena. The Hamiltonian is given by

\begin{equation}
\hat{H}=\frac{k}{8}\left(\hat{N}_{1}-\hat{N}_{2}\right)^{2}-\frac{\mu}{2}\left(\hat{N}_{1}-\hat{N}_{2}\right)-
\frac{\epsilon}{2}\left(\hat{a}_{1}^{\dagger}\hat{a}_{2}+\hat{a}_{2}^{\dagger}\hat{a}_{1}\right)
\label{ham}
\end{equation}
where $\left\{ \hat{a}_{j}^{\dagger},\,\hat{a}_{j}|\, j=1,\,2\right\} $
are the creation and annihilation operators for the condensate $j$, associated respectively to two bosonic Heisenberg 
algebras, with the following commutation relations
\begin{equation}
\left[\hat{a}_{i},\hat{a}_{j}^{\dagger}\right]=\delta_{ij}\;\;\;\;\left[\hat{a}_{i},\hat{a}_{j}\right]=
\left[\hat{a}_{i}^{\dagger},\hat{a}_{j}^{\dagger}\right]=0
\label{heisenberg}
\end{equation}
also the operators $\hat{N}_{j}=\hat{a}_{j}^{\dagger}\hat{a}_{j}$ are the corresponding
boson number operators for each condensate. Since the Hamiltonian does
not depend explicitly on time and commutes with the total boson number
$\hat{N}=\hat{N}_{1}+\hat{N}_{2}$, the total number of bosons $N$ is a conserved quantity
and it is possible to set ourselves to a subspace with fixed value of $N$. The coupling $k$ provides the
strength of the scattering interaction between bosons and may be attractive
$\left(k<0\right)$ or repulsive $\left(k>0\right)$. The parameter
$\mu$ is the external potential which corresponds to an asymmetry
between the condensates and $\epsilon$ is the coupling for the tunneling.
Despite its apparent simplicity, this model exhibits very interesting behaviour. 
In particular, an investigation of its quantum dynamics predicts non-trivial threshold couplings \cite{milb55,Links02}, in 
qualitative agreement with experimental results \cite{albiez}.
We restrict ourselves to study the case $k<0$ because it is known that
in the attractive case the system presents a quantum phase transition
(QPT) \cite{pan,Links02}.

\subsection{Classical analysis}\label{CAbose-hubb}

Let $\left\{ \hat{N}_{j},\,\hat{\theta}_{j}|\, j=1,2\right\} $ be
quantum variables satisfying the canonical commutation relations.
In order to go to the classical limit it is convenient to make a change of variables from the operators $\left\{ \hat{a}_{j}^{\dagger},\,\hat{a}_{j}|\, j=1,\,2\right\} $
to a number-phase representation via 
\begin{equation}
\hat{a}_{j}=e^{i\hat{\theta}_{j}}\sqrt{\hat{N}_{j}},\,\,\hat{a}_{j}^{\dagger}=\sqrt{\hat{N}_{j}}e^{-i\hat{\theta}_{j}},
\end{equation}
such that the Heisenberg canonical commutation relations are preserved.
Now we define the variables
\begin{equation}
\hat{z}=\frac{1}{\hat{N}}\left(\hat{N}_{1}-\hat{N}_{2}\right)
\end{equation}
representing the fractional occupation imbalance and
\begin{equation}
\hat{\theta}=\frac{\hat{N}}{2}\left(\hat{\theta}_{1}-\hat{\theta}_{2}\right)
\end{equation}
representing the phase difference. Note that $\left(\hat{z},\,\hat{\theta}\right)$
are canonically conjugate variables. In the classical limit where $N$
is large, but still finite, we may equivalently consider the Hamiltonian
\cite{PhysRevA.59.620, Links01}

\begin{equation}
H\left(z,\theta\right)=\frac{\epsilon N}{2}\left[\frac{\lambda}{2}z^{2}-\beta z-\sqrt{1-z^{2}}cos\left(\frac{2\theta}{N}\right)\right]\label{classicalH}
\end{equation}
where $\lambda=\frac{kN}{2\epsilon},\,\,\beta=\frac{\mu}{\epsilon}$ are the parameters governing the different 
dynamic regimes of the condensates atomic tunneling.
The Hamilton's equations of motion of the system are given by
\begin{equation}
\dot{\theta}=-\frac{\partial H}{\partial z}=\frac{\epsilon N}{2}\left[\lambda z-\beta+\frac{z}{\sqrt{1-z^{2}}}cos\left(\frac{2\theta}{N}\right)\right],
\end{equation}

\begin{equation}
\dot{z}=\frac{\partial H}{\partial\theta}=-\epsilon\left[\sqrt{1-z^{2}}sin\left(\frac{2\theta}{N}\right)\right].
\end{equation}

\begin{figure}[H]
\centering{}\includegraphics[scale=0.8]{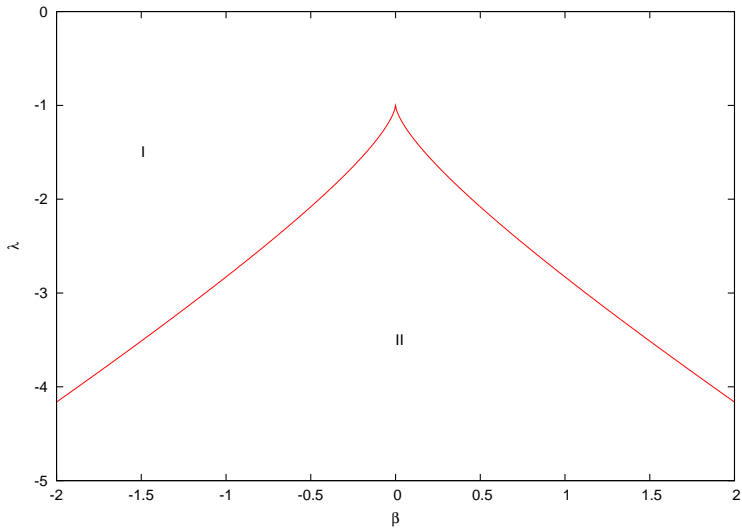}\caption{Coupling parameter space diagram 
identifying the different types of
solutions for the equations of fixed points $\dot{z}=\dot{\theta}=0$.
In the region $I$ there is one solution for $\theta=\frac{N\pi}{2}$
and one solution for $\theta=0$ and in region $II$ there is one
solution for $\theta=\frac{N\pi}{2}$ and three solutions for $\theta=0$.}\label{fig:01}
\end{figure}

\noindent Here we would like to mention that the above classical equations of motion
were also derived in \cite{PhysRevA.59.620}. Now we study the fixed points
of the Hamiltonian (\ref{classicalH}), determined from the above equations by the condition
$\dot{z}=\dot{\theta}=0$. By performing a numerical analysis we find
that these equations may have one, two or three solutions, depending
on the values of the coupling parameters (see \cite{Links03} for details).
This allow us to construct a diagram of parameters identifying the
different types of solutions, depicted in Figure \ref{fig:01}. We remark that
in the absence of the external potential $\left(\beta=0\right)$,
we have a fixed point bifurcation given by $\lambda_{0}=-1$.

It has been conjectured \cite{PhysRevA.71.042303,PhysRevA.65.042107} that fixed points
in the classical analysis can be used to identify quantum phase transitions
in a general level, regardless of the nature of the bifurcation. This
model exhibits a bifurcation point in the parameter diagram and becomes a
natural candidate to study quantum phase transitions.

\subsection{Bethe ansatz solution}

The two-site Bose-Hubbard model has a very special property, it is
an exactly solvable model. The integrability of this model was demonstrated
using the \emph{Quantum Inverse Scattering Method (QISM) }\cite{Links01}.
A major advantage of the QISM, or \emph{algebraic Bethe ansatz method}   
is that it leads to an exact solution for the integrable models, furnishing the energy eigenvalues 
and the Bethe ansatz equations (BAE) for the models. 
The set of BAE and energies for the Hamiltonian (\ref{ham}) can be written in two different
ways: (i) in a product form as well as (ii) in an additive form \cite{Links04}.
Here we will employ  the additive form, given by \cite{Links04}

\begin{equation}
E=\frac{kN^{2}}{8}-\frac{\mu N}{2}+\frac{\epsilon}{2}\sum_{j=1}^{N}\upsilon_{j},
\label{eeB-H}
\end{equation}
where the parameters $\upsilon_{j}$ must satisfy the following \emph{BAE}:

\begin{equation}
\frac{\epsilon\upsilon_{j}^{2}+\left[k\left(1-N\right)-2\mu\right]\upsilon_{j}-\epsilon}{k\upsilon_{j}^{2}}=\sum_{k\neq j}^{N}\frac{2}{\upsilon_{k}-\upsilon_{j}}.
\label{BAE_B-H}
\end{equation}

Thus, each set $\left\{ \upsilon_{j},\, j=1,\,...,\, N\right\} $
solution of BAE (\ref{BAE_B-H}), provides an energy eigenvalue
(\ref{eeB-H}) of the Hamiltonian. In principle, it is not realizable
the analytical solution of these equations. There are, however, a few studies in 
asymptotic limits for the repulsive case \cite{Links05}.
Nevertheless due to their structure, in order to achieve a more
accurate analysis of the model, we may employ numerical tools. The numerical
solution of the equations (\ref{BAE_B-H}), shows that the ground state in
the attractive case has the structure of a \emph{N-string} - i.e., it presents itself
as a collection of symmetric solutions in the complex plane like $\left\{ \nu_{j}\mid j=1,...,N\right\} =\left\{ x_{k}\pm iy_{k}\mid k=1,...,\frac{N}{2}\right\} $
- in contrast with the repulsive case, where the ground-state solution
has always real roots. All these solutions have been checked with
the exact diagonalization of the Hamiltonian and there is a full agreement.
In the following charts we show the solutions of the BAE for some
values of the total number of particles and we adopted $\mu=0$ by the consideration of
two main reasons: $\left(i\right)$ nonzero values of $\mu$ do not
significantly alter the behavior of the system, just shifting the
energy levels \cite{Links02}; $\left(ii\right)$, much of the experimental
realizations with these systems are made on the condition of zero
external potential \cite{PhysRevLett.105.204101}. For further analysis, it is useful
to take into account the parameter $\lambda^{-1}$, borrowed from the classical
analysis:
\begin{equation}
\lambda^{-1}=\frac{2\epsilon}{kN}.
\end{equation}

\begin{figure}[H]
\centering{}\includegraphics[scale=0.8]{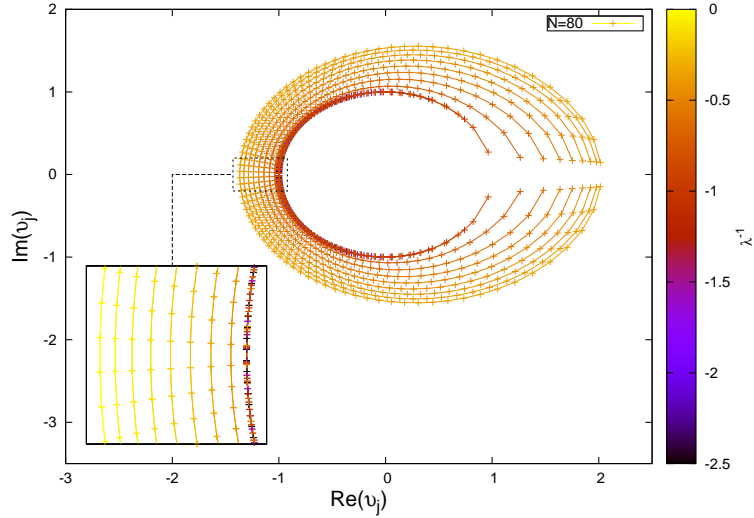}\caption{Solutions of Bethe ansatz 
equations (\ref{BAE_B-H}) in the complex plane for the ground
state considering the particular case $N=80$. Each curve corresponds
to a value of the parameter $\lambda^{-1}$ as the color table, right,
and sits on a set of points (roots of the eq. (\ref{BAE_B-H})).
Here, and in the subsequent figures, the inset shows an enlarged view of the delimited area. The abrupt change in the distribution of roots occurs at
$\lambda^{-1}\approx-0.833$.}
\label{fig:02}
\end{figure}

\begin{figure}[H]
\centering{}\includegraphics[scale=0.8]{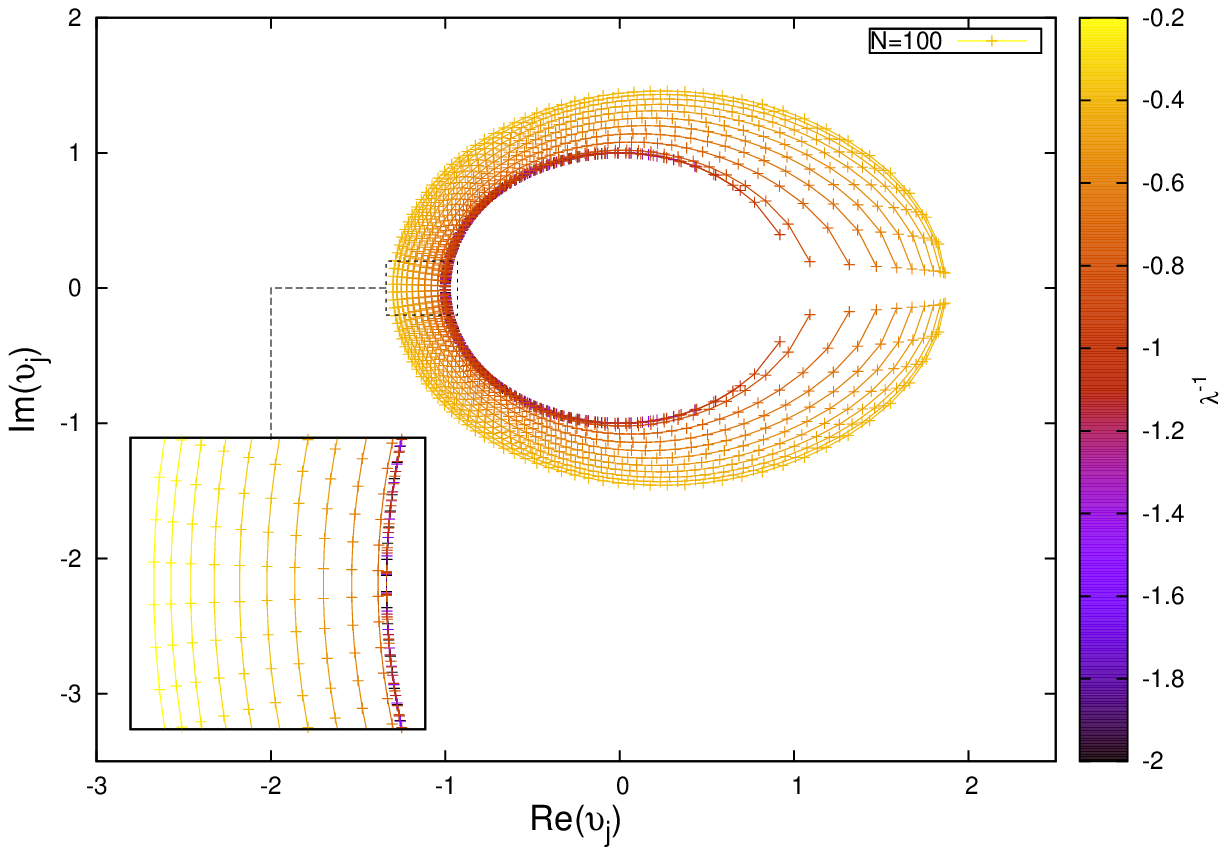}\caption{Solutions of Bethe ansatz 
equations (\ref{BAE_B-H}) in the complex plane 
for the ground
state considering the particular case $N=100$. Each curve corresponds
to a value of the parameter $\lambda^{-1}$ as the color table, right,
and sits on a set of points (roots of the eq. (\ref{BAE_B-H})). The abrupt change in the distribution of roots occurs at
$\lambda^{-1}\approx-0.909$.}
\label{fig:03}
\end{figure}

\begin{figure}[H]
\centering{}\includegraphics[scale=0.8]{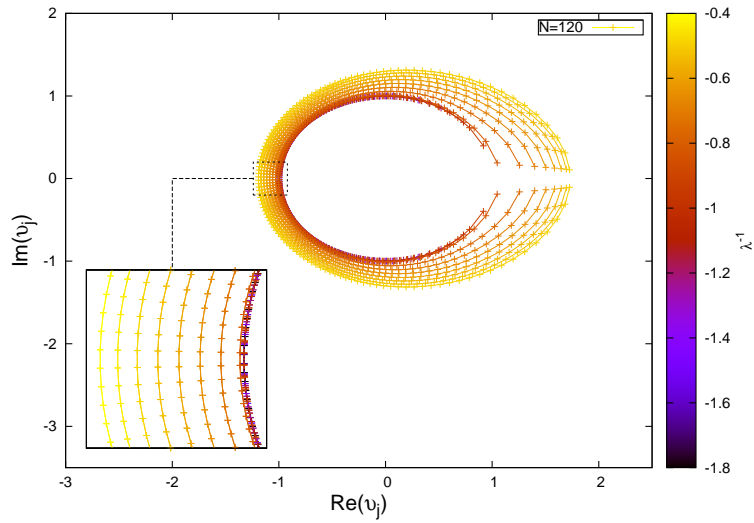}\caption{Solutions of Bethe ansatz 
equations (\ref{BAE_B-H}) in the complex plane 
for the ground
state considering the particular case $N=120$. Each curve corresponds
to a value of the parameter $\lambda^{-1}$ as the color table, right,
and sits on a set of points (roots of the eq. (\ref{BAE_B-H})). The abrupt change in the distribution of roots occurs at
$\lambda^{-1}\approx-0.926$.}
\label{fig:04}
\end{figure}

In Figure \ref{fig:02}, notice the abrupt change in the distribution of roots
of eq. (\ref{BAE_B-H}) in the complex plane from a certain critical value $\lambda_{C}^{-1}$.
For values of $\lambda^{-1}$ larger then $\lambda_{C}^{-1}$ the
roots begin to distribute about distinct families of curves. The same
characteristic behavior of the ground state roots of the Bethe ansatz equations
is observed for other values of $N$ - see Figure \ref{fig:03} and Figure \ref{fig:04}.

The largest value of $N$ for which we have numerically solved the
\emph{BAE} (\ref{BAE_B-H}) and to determine the configurations
of the roots into the ground state was $N=120$, as illustrated in
Figure \ref{fig:04}. It is interesting to note that equations similar to (\ref{BAE_B-H})
were found and solved for other integrable models of interest such
as the reduced BCS model \cite{Roman2002483} and the BCS wave-p type model
\cite{Ibanez:2008xq}, where it was possible to solve the \emph{BAE} for
$N=100$ and $N=30$ Cooper pairs, respectively.

Later we will discuss this peculiar behavior of the solutions of \emph{BAE}
(\ref{BAE_B-H}) under the light of the physical phenomena occurring in the
system.

\subsection{Quantum Phase Transitions}\label{QPT_hubb}

As discussed in the classical analysis, the fixed points can be used
to identify quantum phase transitions in a general level. The \emph{two-site
Bose-Hubbard model} has a bifurcation point that can be seen in the parameters
diagram presented in Figure \ref{fig:01} and becomes a natural candidates for the study of entanglement
and quantum phase transitions. Quantum Phase Transitions (\emph{QPT})
occur at absolute zero temperature due to quantum fluctuations when
we vary some parameter \cite{Sachdev2011}. Although QPT are
rigorously defined only in the thermodynamic limit, when N goes to infinity, the analysis made
in finite systems, using the techniques below, indicates points that show a response to 
variations in the ground state and indicates the presence of quantum phase transitions \cite{PhysRevLett.92.212501}.
Then in order to simplify the nomenclature we use the terminology \emph{QPT} also when referring to these
points.There are different techniques to identify
them and, in particular, the critical values of the parameters for
which the transitions happen. We will use three different ones, all of them sensible to the ground state 
behavior, entanglement, energy gap and fidelity:

\subsubsection{Entanglement}

\begin{figure}[H]
\centering{}\includegraphics[scale=0.8]{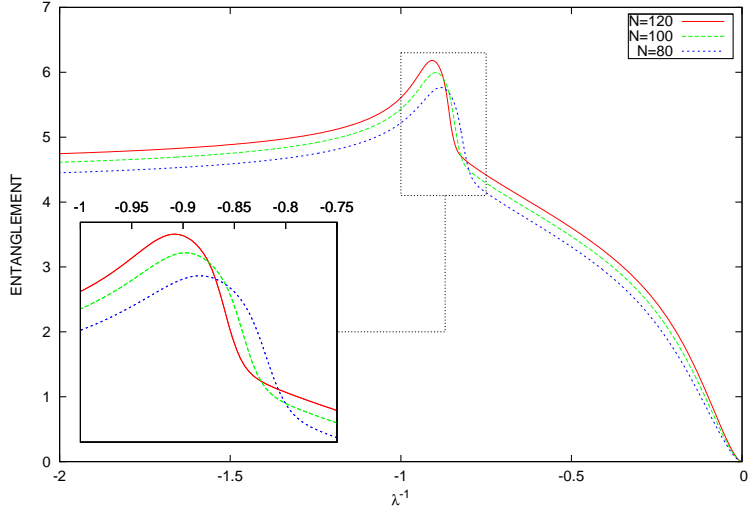}\caption{Entanglement entropy of the ground state as a 
function of $\lambda^{-1}$
for different values of the total number of atoms $N$. The maximum
of the entanglement, indicating the \emph{QPT} point, occurs at $\lambda^{-1}=-0.88,\,-0.91,\,-0.93$
for $N=80,\,100,\,120$, respectively. As $N$ increases the maximum
entropy moves to the critical point $\lambda_{C}^{-1}=-1$.}
\label{fig:05}
\end{figure}

We can consider the pair of coupled Bose-Einstein
condensates as a bipartite system of two modes, \textquotedbl{}1\textquotedbl{}
and \textquotedbl{}2\textquotedbl{}. In this case, the standard measure
of entanglement is the von Neumann entropy of the reduced density
operator of either of the modes \cite{milb55}. The state of each
mode is characterized by its occupation number. Using the fact that
the total number of atoms $N$ is constant, a general state of the
system can be written in terms of the Fock states by
\begin{equation}
|\Psi\rangle=\sum_{n=0}^{N}c_{n}|n\rangle|N-n\rangle
\end{equation}
 where ${c_{n}}$ are complex numbers and for $N_{1}=n$ atoms in
condensate $1$ there will be $N_{2}=N-n$ atoms in condensate $2$.
The density operator of the system is given by 
\begin{equation}
\rho=|\Psi\rangle\langle\Psi|=\sum_{m,n=0}^{N}c_{m}^{*}c_{n}|m\rangle|N-m\rangle\langle n|\langle N-n|.
\end{equation}
Taking the partial trace with respect to mode \textquotedbl{}2\textquotedbl{}
yields the reduced density operator for mode \textquotedbl{}1\textquotedbl{},
\begin{equation}
\rho_{1}=\textrm{Tr}_{2}(\rho)=\sum_{n=0}^{N}|c_{n}|^{2}|n\rangle\langle n|
\end{equation}
Thus the entropy of entanglement of the ground-state of the system
is given by
\begin{equation}
E(\rho_{1})=-\textrm{Tr}[\rho_{1}\textrm{log}(\rho_{1})]=-\sum_{n=0}^{N}|c_{n}|^{2}\textrm{log}(|c_{n}|^{2}).
\label{entanglement}
\end{equation}
The value of the parameter for which the entropy of entanglement has
a maximum identifies the parameters of the \emph{QPT} \cite{PhysRevLett.93.086402}, depicted in  Figure \ref{fig:05}.
We fix the parameter $\epsilon=1$ and use $k$ as the negative variable coupling parameter.

\subsubsection{Energy Gap}

\begin{figure}[H]
\centering{}\includegraphics[scale=0.8]{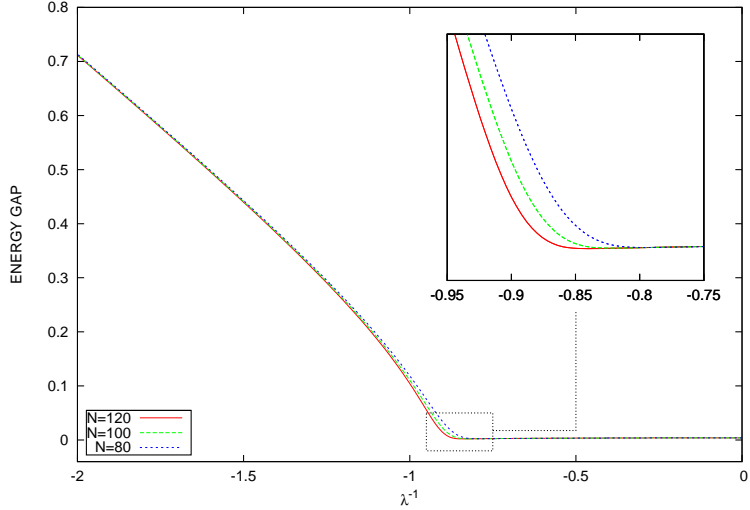}\caption{Energy gap of the ground state as a function 
of $\lambda^{-1}$ for
different values of the total number of atoms $N$. The minimum of
the energy gap, indicating the \emph{QPT} point, occurs at $\lambda^{-1}=-0.80,\,-0.83,\,-0.85$
for $N=80,\,100,\,120$, respectively. As $N$ increases the minimum
of the energy gap moves to the critical point $\lambda_{C}^{-1}=-1$.}
\label{fig:06}
\end{figure}

We consider now the energy gap, which is defined
as the difference between the first excited state and the ground-state
of the system,
\begin{equation}
\Delta E=E^{\left(1\right)}-E^{\left(0\right)}.
\label{gap}
\end{equation}
The value of the parameter for which the gap is zero, or has a minimum, 
identifies the parameters of the QPT \cite{Sachdev2011}. 
In Figure \ref{fig:06} we present the energy gap as a function of $\lambda^{-1}$ for 
$N=80, 100$ and $120$.

\subsubsection{Fidelity}

Another possibility to investigate the QPT is through
the behavior of the fidelity, which is a concept widely used in the
Quantum Information Theory \cite{nielsen00}. The fidelity is basically
defined as the modulus of the wavefunction overlap between two states
\begin{equation}
\mathcal{F}\left(\psi,\phi\right)=\left|\left\langle \psi|\phi\right\rangle \right|
\label{fidelity}
\end{equation}
Basically, the point where the fidelity has a sharp decline defines
a critical value for the parameter \cite{PhysRevLett.98.110601}.

\begin{figure}[H]
\centering{}\includegraphics[scale=0.8]{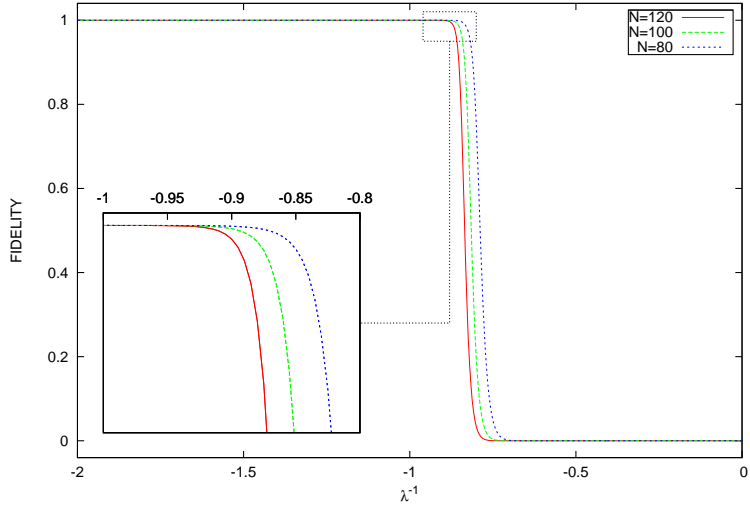}\caption{Fidelity of the ground state as a function 
of $\lambda^{-1}$ for
different values of the total number of atoms $N$. The value at which
the fidelity has an abrupt decay, indicating the \emph{QPT} point,
occurs at $\lambda^{-1}=-0.88,\,-0.91,\,-0.93$ for $N=80,\,100,\,120$,
respectively. As $N$ increases the sharp
decline moves to the critical point $\lambda_{C}^{-1}=-1$.}
\label{fig:07}
\end{figure}

\subsection{BAE and QPT}
Observing the graphs - Figure \ref{fig:05}, Figure \ref{fig:06} and Figure \ref{fig:07} - we
identify the critical points of \emph{QPT }for a given set of parameters
and we note the striking correspondence between the points obtained by investigating the
entanglement, energy gap and fidelity, that indicate a \emph{QPT}, and the point
where the respective ground state solutions of the \emph{BAE} change their behavior.
We also note that, as the number of particles increases, the
correlations between the transition point predicted by standard methods
and the one predicted by the change of behavior of solutions of \emph{BAE
} agrees with greater precision and in all cases it tends the value
$\lambda_{C}^{-1}=-1$, as predicted by the classical analysis. Therefore the behavior of the ground state of the model
translated as the behavior of the set of solutions of the \emph{BAE }, when the relevant parameters are changed, can 
be used as an alternative
method to locate the points of the phase transition of the system.\\
 
\begin{figure}[H]
\centering{}\includegraphics[scale=0.8]{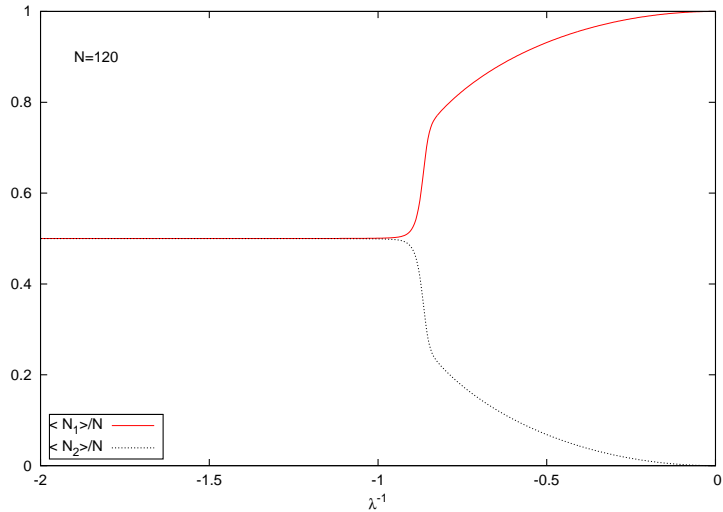}\caption{
Ground state expectation values for the fraction of the number of atoms in the condensates
$1$ and $2$ as a function of $\lambda^{-1}$. We can observe an abrupt change in the behavior 
near the critical point $\lambda_{C}^{-1}=-1$
}
\label{fig:08}
\end{figure}

We interpret this QPT through the behavior of
the expected value of the normalized number of atoms in the condensates.
In Figure \ref{fig:08} we show the ground state expectation value of the
normalized number of particles in the condensates $1$ and $2$, for $N=120$ particles.
We observe that for low values of $\lambda^{-1}$, the expected value of the normalized number of
atoms in the condensates $1$ and  $2$ is $\frac{\left\langle N_{1}\right\rangle }{N}\cong0.5$ and 
$\frac{\left\langle N_{2}\right\rangle }{N}\cong0.5$, respectively, 
meaning that approximately half of the atoms is in the condensate
$1$ and the other half is in the condensate $2$. By varying the parameter $\lambda^{-1}$
we find that this value suffers a sharp drop near the critical point $\lambda_{C}=-1$,
approaching the values $\frac{\left\langle N_{1}\right\rangle }{N}\cong1.0$ and 
$\frac{\left\langle N_{2}\right\rangle }{N}\cong0.0$, respectively, 
after crossing this point. So we can interpret this transition as
the separation between a \emph{delocalized phase}, with tunneling
of atoms between two condensates, and a \emph{localized phase}. 

\section{Hetero-atomic molecular Bose-Einstein condensate}

We now consider a model for two distinct species of atoms that form a hetero-nuclear molecular Bose-Einstein 
condensate and is described by a three mode Hamiltonian. A novel feature of a molecular Bose-Einstein 
condensate is that atomic and molecular states can exist as a superposition \cite{PhysRevA.78.061601}.
For the cases where the molecules are heteronuclear, the presence of a
permanent electric dipole moment also opens the possibility for manipulating
the condensate through electrostatic forces \cite{PhysRevA.70.021402}. These characteristics make the 
study of such models increasingly interesting allowing to access phenomena that goes from novel quantum 
phase transitions to ultracold chemistry \cite{1367-2630-11-5-055049}.The Hamiltonian
reads
\begin{equation}
\hat{H}=\mu_{c}\hat{N}_{c}+\Omega\left(\hat{a}^{\dagger}\hat{b}^{\dagger}\hat{c}+\hat{c}^{\dagger}\hat{a}\hat{b}\right)
\label{Habc},
\end{equation}
where we label each atomic species by $a$ and $b$, which can combine to produce a molecule labelled
by $c$. 
This is the Hamiltonian studied in \cite{PhysRevA.1.446}
and also known as the \emph{optical Hamiltonian}. Again $\left\{ \hat{j},\,\hat{j}^{\dagger}|\, j=a,\, b,\, c\right\} $
are the canonical creation and annihilation operators satisfying the
usual bosonic Heisenberg commutation relations of eq.(\ref{heisenberg}) and $\hat{N}_{j}=\hat{j}^{\dagger}\hat{j}$
are the corresponding boson number operators. The parameter $\mu_{c}$
is the external potential and $\Omega$ is the amplitude for interconversion
of atoms and molecules. The Hamiltonian commutes with $\hat{J}=\hat{N}_{a}-\hat{N}_{b}$
and with the total number $\hat{N}=\hat{N}_{a}+\hat{N}_{b}+2\hat{N}_{c}$.
We refer to $\hat{J}$ as the \emph{atomic imbalance} and introduce
$k=\frac{J}{N},\, k\,\in\left[-1,\,1\right]$, as the \emph{fractional
atomic imbalance.} A discussion of the physical implications of the imbalance can be 
found, for example, in \cite{mattei, pu5, weiping}.

\subsection{Classical analysis}

Let $\left\{ \hat{N}_{j},\,\hat{\theta}_{j}|\, j=a,\, b,\, c\right\} $
be quantum variables satisfying the canonical commutation relations
\begin{equation}
\left[\hat{N}_{j},\hat{\theta}_{k}\right]=i\delta_{jk}I,\,\,\left[\hat{N}_{j},\hat{N}_{k}\right]=\left[\hat{\theta}_{j},\hat{\theta}_{k}\right]=0.
\end{equation}
As in section \emph{\ref{CAbose-hubb}} we make a change of variables from the operators $\hat{j}^{\dagger},\,\hat{j}$
to a number-phase representation via
\begin{equation}
\hat{j}=e^{i\hat{\theta}_{j}}\sqrt{\hat{N}_{j}},\,\,\hat{j}^{\dagger}=\sqrt{\hat{N}_{j}}e^{-i\hat{\theta}_{j}},
\end{equation}
such that the Heisenberg canonical commutation relations are preserved.
Now we define the canonically conjugate variables $\hat{z}$ and $\hat{\theta}$
as 
\begin{equation}
\hat{z}=\frac{1}{\hat{N}}\left(\hat{N}_{a}+\hat{N}_{b}-2\hat{N}_{c}\right),
\end{equation}
\begin{equation}
\hat{\theta}=\frac{\hat{N}}{4}\left(\hat{\theta}_{a}+\hat{\theta}_{b}-\hat{\theta}_{c}\right).
\end{equation}
In the limit of large $N$ we can obtain from the \emph{optical Hamiltonian}
the approximate (rescaled) classical Hamiltonian
\begin{equation}
H\left(z,\theta\right)=2\alpha\left(z-1\right)+\sqrt{2\left(1-z\right)\left(z+c_{+}\right)\left(z+c_{-}\right)}cos\left(\frac{4\theta}{N}\right)
\label{opticalH}
\end{equation}
where $\alpha=-\frac{\mu_{c}}{\Omega\sqrt{2N}}$, and $c_{\pm}=1\pm2k$.

The Hamilton's equations of motion are given by
\begin{equation}
\dot{\theta}=-\frac{\partial H}{\partial z}=2\alpha+\frac{2\left(1-z\right)\left(1+z\right)-\left(z+c_{+}\right)\left(z-c_{-}\right)}{\sqrt{2\left(1-z\right)\left(z+c_{+}\right)\left(z+c_{-}\right)}}cos\left(\frac{4\theta}{N}\right),
\end{equation}

\begin{equation}
\dot{z}=\frac{\partial H}{\partial\theta}=-\frac{4}{N}\sqrt{2\left(1-z\right)\left(z+c_{+}\right)\left(z+c_{-}\right)}sin\left(\frac{2\theta}{N}\right).
\end{equation}
From these equations, the fixed points of the system can be determined by the condition $\dot{z}=\dot{\theta}=0$.
It was shown in \cite{mattei} that this model exhibits a \emph{QPT} just in the case of zero imbalance, ie, $k=0$. 
For this reason we deal here only with this case, for which it was demonstrated that 
there are two fixed points $\alpha=1$ and $\alpha=-1$, associated to a minimum and to a maximum in the parameter
phase diagram, respectively \cite{mattei}.

Therefore, for the quantized system, it is expected that $\alpha=1$ may indicate a
\emph{QPT}.

\subsection{Bethe ansatz solution}

The Hamiltonian \ref{Habc} is also an integrable model \cite{Links01}.
The energy eigenvalues of the hetero-atomic molecular Bose-Einstein
condensates are given by

\begin{equation}
E=-\Omega\sum_{j=1}^{\frac{\left(N-J\right)}{2}}\nu_{j}
\label{ee_Hetero}
\end{equation}
where the parameters $\nu_{j}$ must satisfy the following Bethe ansatz
equations (BAE):

\begin{equation}
\frac{J+1}{\nu_{j}}-\nu_{j}-\frac{\mu}{\Omega}=\sum_{k\neq j}^{\frac{\left(N-J\right)}{2}}\frac{2}{\nu_{k}-\nu_{j}}.
\label{BAE_Hetero}
\end{equation}

The numerical solution of the equations (\ref{BAE_Hetero}) shows that the ground
state has the structure of real roots solely. All these solutions have been
checked with the exact diagonalization of the Hamiltonian and there
is a full agreement.

\begin{figure}[H]

\centering{}\includegraphics[angle=0.0, scale=0.8]{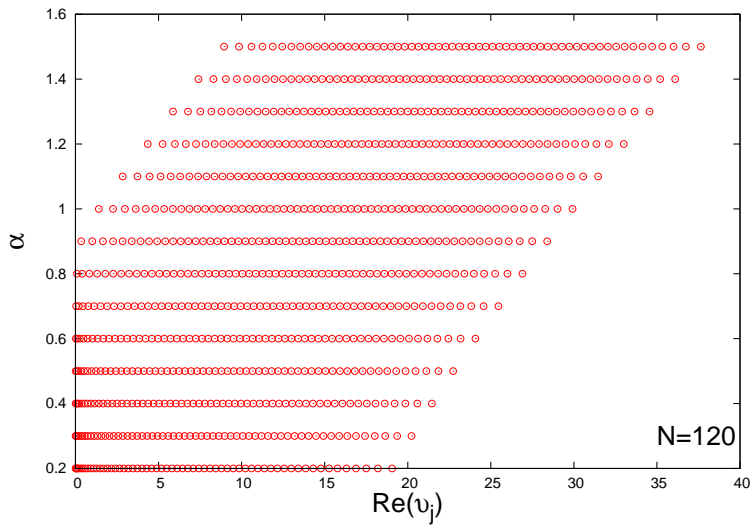}\caption{Solutions of Bethe ansatz 
equations (\ref{BAE_Hetero}) for the ground state versus $\alpha$ considering the case $N=120$.
Each line corresponds to a BA-solution for a particular value of the parameter $\alpha$. 
There is a change in the behaviour of the distribution of the BA-roots around the critical value 
$\alpha_{C}=1$
}
\label{fig:3.1}
\end{figure}

Notice the abrupt change in the distribution of roots of eq. (\ref{BAE_Hetero})
from a certain critical value $\alpha_{C}$, shown in \ref{fig:3.1}. For values of $\alpha$
smaller than $\alpha_{C}$ the roots are tightly packed when close to the origin
and we have a distribution beginning exactly at the origin. By increasing
the value of the parameter $\alpha$ the density of roots are not
so dense at the origin and tend to get distanced from it. This process
reaches its culmination in $\alpha_{C}$ approximately $1$, where
no roots are at zero. After $\alpha_{C}$ the set of roots moves away
from the origin. Therefore, a different behavior of the \emph{BAE}-solutions when we cross the 
region $\alpha<1$ to the region $\alpha>1$ is observed.
For other values of $N$, such as
$N=80$ and $N=100$, a very similar behaviour of the ground state BA-roots is also obtained.
In the case of non-zero imbalance $\left(k\neq0\right)$, all
roots are away from the origin, independent of the value of $\alpha$.
In this case, where it is known that there is no \emph{QPT,} we also
do not observe a different behavior in the solution of the \emph{BAE}.

\subsection{Quantum Phase Transitions}
Here as in section \emph{\ref{QPT_hubb}} we will use three different techniques to identify quantum phase 
transitions: entanglement, energy gap and fidelity. We use the same definitions as in 
eqs.(\ref{entanglement}-\ref{fidelity}) and the results are shown in the next figures (\ref{fig:3.2}-\ref{fig:3.4}).

\begin{description}
\item[{Entanglement}]~
\end{description}

\begin{figure}[H]
\centering{}\includegraphics[angle=-90, scale=0.4]{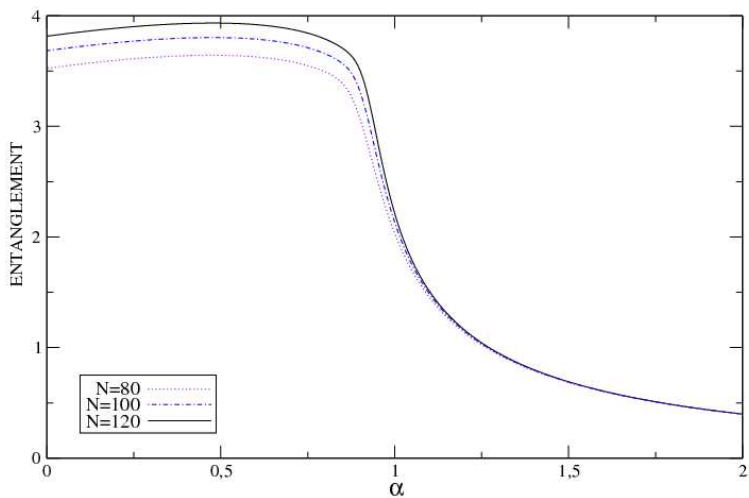}\caption{Entanglement entropy of the ground state as a 
function of $\alpha$
for different values of the total number of atoms $N$. The entanglement entropy exhibits a sudden decrease close to the critical point $\alpha_c =1$ that
becomes more pronounced as $N$ increases.}
\label{fig:3.2}
\end{figure}

\begin{description}
\item [{Energy~Gap}]~
\end{description}

\begin{figure}[H]
\centering{}\includegraphics[angle=-90, scale=0.4]{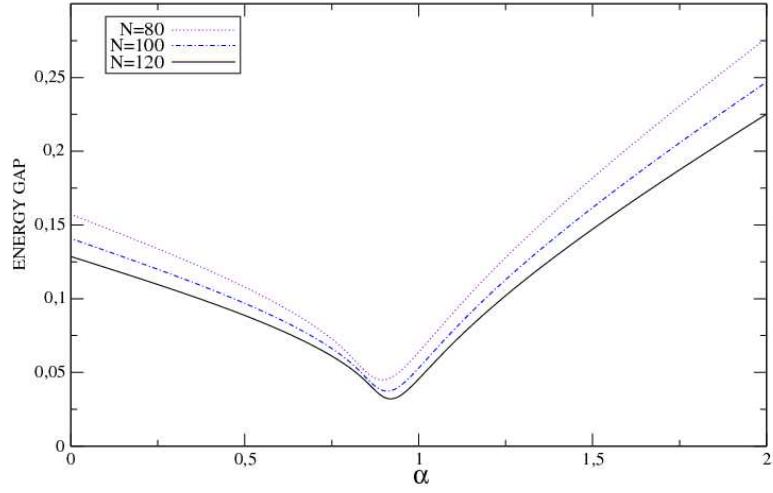}\caption{Energy gap of the ground state as a function of $\alpha$ 
for different
values of the total number of atoms $N$. As $N$ increases the minimum
of the energy gap moves to the critical point $\alpha_{C}=1$.}
\label{fig:3.3}
\end{figure}

\begin{description}
\item [{Fidelity}]~
\end{description}

\begin{figure}[H]
\centering{}\includegraphics[angle=-90, scale=0.4]{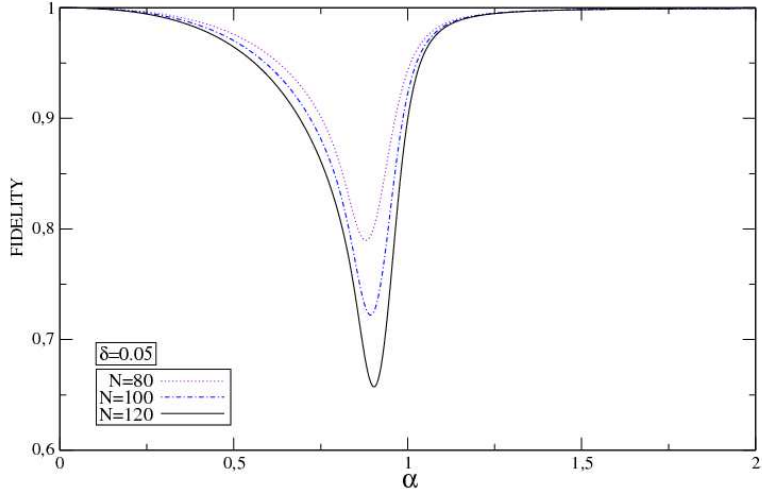}\caption{Fidelity of the ground state as a function of $\alpha$ for different
values of the total number of atoms $N$. As $N$ increases the minimum moves to the critical point $\alpha_{C}=1$.}
\label{fig:3.4}
\end{figure}

\subsection{BAE and QPT}

Again, we find an agreement between the point $\alpha_{C}=1$ where
the behavior of the solutions of
the \emph{BAE } sharply changes and the critical point obtained from the other techniques. 
So in the same way as for the \emph{Bose-Hubbard model}, we are led to the conclusion that
the solutions of the \emph{BAE} can also be used as an alternative
method for the identification of \emph{QPT}. \\

\begin{figure}[H]

\centering{}\includegraphics[scale=0.4]{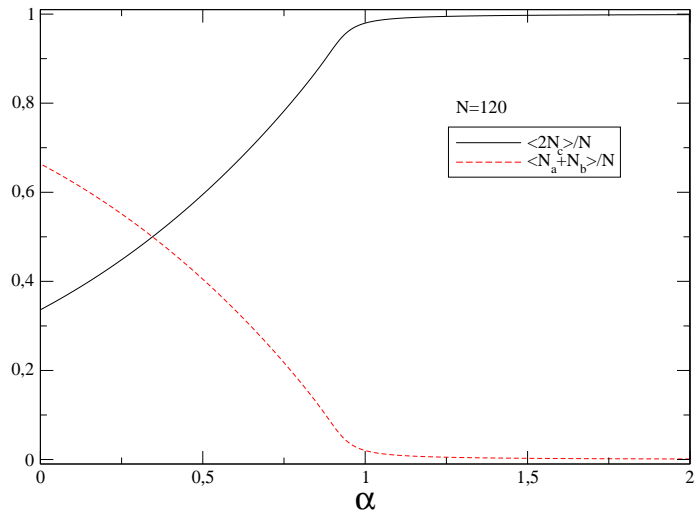}\caption{
Normalized ground state expectation value of the number of atoms 
and molecules 
in the condensate as a function of the parameter $\alpha$. We can observe a sharp transition close to the critical point $\alpha_C=1$.
}
\label{fig:3.5}
\end{figure}

We can see in Figure \ref{fig:3.5} the correlation between the saturation of
the number of molecules and the critical point $\alpha_{C}=1$. We note
that the ground state expectation value of the normalized number of molecules increases almost linearly with
increasing value of $\alpha$. When the normalized number of molecules reaches
the saturation point 

we identify the critical
point of the system. Therefore, this point can be interpreted as the critical
value for which there are just molecules in the system.

\section{Summary}

The main purpose of this study was to demonstrate how the presence
of very significant physical phenomena can also be inferred from the
Bethe ansatz structure of a given integrable model. To illustrate this
idea, we investigated the two-site Bose-Hubbard model and the 
hetero-atomic molecular Bose-Einstein condensate, two well known exactly solvable models,
and explored their physical and mathematical properties. We began
with a classical analysis for each of models exploiting the fixed point 
structure and making explicit the presence of bifurcation points for 
critical values of the relevant parameters. We then presented their energy
eigenvalues by means of the quantum inverse scattering method. The
\emph{BAE}'s thus obtained are quite involved and, apart from 
some some limiting situations, it is virtually unfeasible to obtain an analytical solution. 
Nevertheless the structure of the \emph{BAE}'s for each model allows the 
possibility of obtaining well behaved numerical solutions. 

The structure of these solutions, although very different for both models, 
present a peculiar behavior when some parameters of the Hamiltonian are varied, indicating that the
ground state solutions of the \emph{BAE}'s are reflecting some change in the system energy spectrum behavior.
It is known that the models in question may experience a \emph{QPT} and the critical
parameters are well known. So, we compared our results coming from the critical points indicated 
by the study of the solution of the \emph{BAE}'s with other methods. We have performed studies 
of the entanglement, energy gap and fidelity for both models and the trend unveiled by these 
methods is highly compatible with the ground state results coming from the analysis of the solutions 
of the \emph{BAE}'s.  

We have found that the quantum phase transitions for the two-site Bose-Hubbard model and for the
hetero-atomic molecular Bose-Einstein condensate have very different interpretations. For the first model the transition reflects a 
separation between a \emph{delocalized} and a \emph{localized} phase, while for the second model the 
transition occurs when there is a saturation of molecules in the system. In particular, 
for the two-site Bose-Hubbard model in the delocalized phase the root distribution is largely independent 
of the coupling, while in the localized phase the roots lie on different arcs. In the case of 
the hetero-atomic molecular Bose-Einstein condensate model all the roots are always on the positive real axis: in the phase with coexistence of  
atoms and molecules the roots distribution begins exactly at the origin, while in the phase where there are only molecules all roots are 
away from the origin. Correspondingly the profiles for both models when 
investigating the entanglement, energy gap and the fidelity are quite different as can be seen directly by 
comparing figures \ref{fig:05} and \ref{fig:3.2}, for the entanglement, \ref{fig:06} and \ref{fig:3.3}, 
for the energy gap, as well as \ref{fig:07} and \ref{fig:3.4}, for the fidelity. 
This suggests that the quantum phase transition for each model is
intrinsically different.

The models that we have analyzed were chosen for their simplicity as well as their physical relevance in the BEC context.
We foresee the possibility of applying this kind of analysis in many different
integrable models and this could possibly lead to some grouping according to the 
geometrical patterns formed by the roots such as arcs and lines, or eventually closed curves in other situations.
Our approach will be particularly convenient in those cases where the exact diagonalization of the model is demanding or when the standard
methods used to identify a \emph{QPT} are not easily implemented.

In any case we firmly believe that the behavior shown by the solutions
of the \emph{BAE} can be used as an alternative method to identify the presence of a \emph{QPT}.

\section*{Acknowledgments }
A. F. thanks J. Links for useful discussions.
I. Roditi acknowledges FAPERJ (Funda\c{c}\~ao Carlos Chagas Filho de Amparo \`a Pesquisa do Estado
do Rio de Janeiro). The authors also acknowledge support from  CNPq (Conselho Nacional de Desenvolvimento Cient\'ifico e Tecnol\'ogico).

\end{document}